\documentclass[aip,jcp,reprint,noshowkeys,superscriptaddress]{revtex4-2}
\makeatletter
\def\switch@array{}
\makeatother

\makeatletter
\let\newfloat\newfloat@dummy
\makeatother

\usepackage{graphicx,dcolumn,bm,xcolor,microtype,multirow,amscd,amsmath,amssymb,amsfonts,physics,wrapfig,bbold,siunitx,xspace,braket,txfonts}
\usepackage[version=4]{mhchem}

\usepackage{placeins}
\usepackage{algorithm}
\usepackage{algpseudocode}
\usepackage{tabularx}
\usepackage{tabularray, lipsum}
\usepackage{booktabs} % For professional rules
\usepackage[utf8]{inputenc}
\usepackage[T1]{fontenc}
\usepackage{longtable}
\usepackage{rotating}
\usepackage{hyperref}
\usepackage{xr}

\hypersetup{
    colorlinks,
    linkcolor={red!50!black},
    citecolor={red!70!black},
    urlcolor={red!80!black}
}

\usepackage{listings}
\definecolor{codegreen}{rgb}{0.58,0.4,0.2}
\definecolor{codegray}{rgb}{0.5,0.5,0.5}
\definecolor{codepurple}{rgb}{0.25,0.35,0.55}
\definecolor{codeblue}{rgb}{0.30,0.60,0.8}
\definecolor{backcolour}{rgb}{0.98,0.98,0.98}
\definecolor{mygray}{rgb}{0.5,0.5,0.5}

\definecolor{sqred}{rgb}{0.85,0.1,0.1}
\definecolor{sqgreen}{rgb}{0.25,0.65,0.15}
\definecolor{sqorange}{rgb}{0.90,0.50,0.15}
\definecolor{sqblue}{rgb}{0.10,0.3,0.60}

\lstdefinestyle{mystyle}{
    backgroundcolor=\color{backcolour},
    commentstyle=\color{codegreen},
    keywordstyle=\color{codeblue},
    numberstyle=\tiny\color{codegray},
    stringstyle=\color{codepurple},
    basicstyle=\ttfamily\footnotesize,
    breakatwhitespace=false,
    breaklines=true,
    captionpos=b,
    keepspaces=true,
    numbers=left,
    numbersep=5pt,
    numberstyle=\ttfamily\tiny\color{mygray},
    showspaces=false,
    showstringspaces=false,
    showtabs=false,
    tabsize=2
  }

  \newcolumntype{d}{D{.}{.}{-1}}

\usepackage{accents}

\newif\ifproduction

\productionfalse

\ifproduction
    \newcommand{\jt}[1]{}
    \newcommand{\np}[1]{}
\else
    \newcommand{\jt}[1]{\textcolor{orange}{#1}}
    \newcommand{\np}[1]{\textcolor{violet}{#1}}
\fi

\newcommand{\bigO}{\mathcal{O}}

\newcommand{\stkout}[1]{\ifmmode\text{\sout{\ensuremath{#1}}}\else\sout{#1}\fi}

\newcommand{\cint}[2]{\left(#1 \,\middle|\, #2 \right)}
\newcommand{\eriao}{\cint{\mu \nu}{\lambda \sigma}}

\newcommand{\hho}{H$_2$O}
\newcommand{\hhon}{(H$_2$O)$_n$}

\newcommand{\UnHam}{Department of Chemistry, University of Hamburg, 22761 Hamburg, Germany; The Hamburg Centre for Ultrafast Imaging (CUI), Hamburg 22761, Germany}

\begin{document}	

\title{Benchmarking Least-Squares Tensor Hypercontraction Techniques for Molecular Systems}

\author{Niklas Paulicks}
		\email{niklas.paulicks@uni-hamburg.de}
		\affiliation{\UnHam}

\author{Johannes T\"olle}
        \email{johannes.toelle@uni-hamburg.de}
        \affiliation{\UnHam}

\newsavebox{\ulbox}
\newcommand{\ulstate}[2][sqred]{%
  \sbox{\ulbox}{#2}%
  \usebox{\ulbox}\llap{\textcolor{#1}{\rule[-0.5ex]{\wd\ulbox}{.8pt}}}%
}

\newcommand{\hlbgstate}[2][sqred!20]{%
  \State \setlength{\fboxsep}{1.5pt}\colorbox{#1}{#2}%
}

% --- Custom Zero-Dependency Algorithm Counter for REVTeX ---
\newcounter{algfigure}
\newcommand{\algorithmname}{Algorithm}

\makeatletter
\newcommand{\fnum@algfigure}{\algorithmname~\thealgfigure}
\newcommand{\ext@algfigure}{lof} % <--- FIX: Redirects list entries to the List of Figures file

\newenvironment{algfigure}{%
  \begin{figure}%
  \def\@captype{algfigure}% Switch counter and label context natively
}{%
  \end{figure}%
}
% Two-column version for reprint format
\newenvironment{algfigure*}{%
  \begin{figure*}%
  \def\@captype{algfigure}%
}{%
  \end{figure*}%
}
\makeatother% Custom square/highlight colors

\definecolor{sqred}{HTML}{D91A1A}
\definecolor{sqgreen}{HTML}{40A626}
\definecolor{sqorange}{HTML}{E68026}
\definecolor{sqblue}{HTML}{1A4C99}

\newcommand{\aux}[1]{\tilde{#1}}

\begin{abstract}
Finding a low-rank approximation for the two-electron integral (ERI) tensor is a crucial step towards reducing the computational cost of many-body electronic structure methods.
In recent years, a range of new tensor factorization techniques, like the so-called tensor-hypercontraction (THC) approach, have been developed to achieve this goal.
Unfortunately, a systematic comparison of accuracy and efficiency of different THC construction algorithms is missing.
In this work, we address this gap by benchmarking these techniques across a wide range of molecular systems.
In addition to offering useful insights into the advantages and disadvantages of the different techniques, we provide reference implementations in the newly developed open-source library, \texttt{PyTHC}.

%\bigskip
%\begin{center}
%	\boxed{\includegraphics[width=0.5\linewidth]{TOC}}
%\end{center}
%\bigskip
\end{abstract}

\maketitle

%%%%%%%%%%%%%%%%%%%%%%%%%%%%%%%%%%%%%%%%%%
\section{Introduction}

The accurate prediction of electronic properties of molecular or material systems is of central interest in computational quantum chemistry and theoretical condensed matter physics. 
Unfortunately, the computational complexity for determining such properties grows rapidly with the electronic degrees of freedom of the system of interest, denoted hereafter by $N$.
For example, Møller--Plesset perturbation theory (MP$n$, where $n$ denotes the perturbation order),\cite{Moller1934NoteApproximation}
or Coupled Cluster theories (CC$n$),\cite{Cizek1966CorrelationProblem} in their canonical form, exhibit
fast-growing complexities of $\mathcal{O}(N^{n+3})$ and
$\mathcal{O}(N^{2n+2})$, respectively.
Thus, the application of most wavefunction methods remains limited to small system sizes and/or small basis sets.

Most of this complexity originates from tensor contractions involving the Electron Repulsion 
Integrals (ERIs), which can be written in Mulliken notation as \cite{Szabo1996ModernQuantum}
\begin{align}
    &\eriao = \int d \mathbf{r}_1 \int d \mathbf{r}_2   \frac{\chi_{\mu} 
    \left(\mathbf{r}_1\right) \chi_{\nu} \left(\mathbf{r}_1\right)\chi_{\lambda} \left(\mathbf{r}_2\right) \chi_{\sigma} \left(\mathbf{r}_2\right)}{| \mathbf{r}_1 - \mathbf{r}_2 |},
\end{align}
where $\chi_{\mu}$ are the spatial basis functions represented by real-valued atom-centered Gaussian-type orbitals (GTOs) throughout this work.
Handling this order-4 tensor implies, without additional approximations, simplifications or symmetry arguments, $\mathcal{O}(N^4)$ costs in storage and at least $\mathcal{O}(N^4)$ computational complexity when contracted with two-index quantities such as the one-particle density matrix in Hartree--Fock (HF) theory. 

Reducing the complexity introduced by the ERI has therefore been a
topic of study for many decades. 
One of the most successful approaches is the Density Fitting / Resolution of the Identity (DF/RI) approximation,
\cite{Whitten1973CoulombicPotential,Beebe1977SimplificationsGeneration,Baerends1973SelfconsistentMolecular,Dunlap1977ApplicabilityLCAOXa,Vahtras1993IntegralApproximations}
where the Coulomb operator $\left(r_{12}\right)^{- 1}$ is projected onto an
auxiliary basis set $P_{\aux{A}}(\mathbf{r})$ which grows linearly with $N$. 
The approximation of the ERI tensor becomes
\begin{align}
    \eriao \approx \sum_{\aux{A} \aux{B}} \cint{\mu \nu}{\aux{A}} \left[ \mathbf{V}^{- 1}\right]_{\aux{A}\aux{B}} \cint{\aux{B}}{\lambda \sigma},
\end{align}
where $\mathbf{V}^{- 1}$ denotes the inverse of the Coulomb integral between auxiliary basis
functions $P_{A}(\mathbf{r}_1)$ and $P_{\aux{B}}(\mathbf{r}_2)$, $\cint{\aux{A}}{\aux{B}}$.
$\mathbf{V}$ is also often referred to as the Coulomb metric.
In this formulation the computational cost is reduced to
$\mathcal{O}(N^3)$ for calculating and storing the $\cint{\mu \nu}{\aux{A}}$ tensor.
The minimal contraction cost is also reduced to $\mathcal{O}(N^3)$.
A similar separable version of RI can equivalently be derived from the Cholesky Decomposition
(CD) of the ERI.\cite{Beebe1977SimplificationsGeneration,Koch2003ReducedScaling,Pedersen2023VersatilityCholesky}
RI has been applied to Kohn--Sham Density Functional Theory (KS-DFT),\cite{Dunlap2000RobustVariational} HF, \cite{Vahtras1993IntegralApproximations} and almost all post-HF methods, see, e.g.,~Refs.~\citenum{Weigend1998RIMP2Optimized,Feyereisen1993UseApproximate,Ishikawa2018RIMP3Calculations,Datta2021MassivelyParallel,Hattig2000CC2Excitation,DePrince2013AccuracyEfficiency,Peng2019CoupledclusterSingles}.

Other decomposition methods
like the Pseudospectral methods (PS), \cite{Friesner1985SolutionSelfconsistent,Friesner1986SolutionHartree,Friesner1987SolutionHartree} the Chain of Spheres Exchange
(COSX) algorithm, \cite{Neese2009EfficientApproximate,Helmich-Paris2021ImprovedChain,Izsak2011OverlapFitted,Neese2009EfficientApproximate} Tensor
Hypercontraction
(THC)\cite{Lu2015CompressionElectron,Parrish2013TensorHypercontraction,Parrish2013DiscreteVariable,Parrish2012TensorHypercontraction,Hohenstein2012CommunicationTensor,Hohenstein2012TensorHypercontraction}
and the Canonical Polyadic Decomposition (CPD)
\cite{Hitchcock1927ExpressionTensor,Lewis2016ClusteredLowRank,Pierce2023EfficientConstruction,Hummel2017LowRank,Pierce2021RobustApproximation,Benedikt2011TensorDecomposition,Benedikt2013TensorDecomposition,Bohm2016TensorRepresentation,Pierce2025UsingMatrixfree}
were proposed to enable further scaling reductions in, e.g., exchange-like
contractions.

Famously, Refs.~\citenum{Hohenstein2012TensorHypercontraction,Parrish2012TensorHypercontraction} introduced the THC factorization of the ERIs into five order-2 tensors (i.e., matrices $\mathbf{X}$ and $\mathbf{Z}$), inspired by the PS family of methods as
\begin{align} 
    \eriao \approx
    \sum_{P Q} X_{\mu}^P X_{\nu}^P Z^{P Q} X_{\lambda}^Q X_{\sigma}^Q.
    \label{eq:thc_decomposition}
\end{align}
As the THC form only involves two-dimensional tensors (i.e., matrices) it can yield
low-scaling tensor contraction paths for many electronic structure methods.
Based on this, low-scaling MP$n$ variants,
\cite{KokkilaSchumacher2015TensorHypercontraction,Hohenstein2012TensorHypercontraction,Parrish2012TensorHypercontraction,Matthews2020ImprovedGrid,Lee2020SystematicallyImprovable,Matthews2021CriticalAnalysis,Zhang2025BlockTensor}
self-consistent field (SCF) methods,
\cite{Hu2017InterpolativeSeparable,Hillers-Bendtsen2025AcceleratingHartree,Hillers-Bendtsen2025LoweringScaling,Lee2020SystematicallyImprovable,Li2026DeterministicTensor}
CC energy contributions
\cite{Datar2024RobustTensor,Hohenstein2022RankreducedCoupledcluster,Schutski2017TensorStructuredCoupled}
CCSD with transcorrelated Hamiltonians,\cite{Liao2026InterpolativeSeparable}
as well as Quantum Embedding,\cite{Yang2026InitioManybody} $GW$ and Random Phase Approximation
(RPA)  \cite{Duchemin2021CubicScalingAllElectron,Duchemin2019SeparableResolutionoftheidentity,Pokhilko2024TensorHypercontraction,Qin2023InterpolativeSeparable,Yin2025SpatialSignatures,Yeh2023LowScalingAlgorithm} calculations have been developed.

In Ref.~\citenum{Hohenstein2012TensorHypercontraction}, the different THC
matrices were determined by CPD using a non-linear
optimization procedure. In order to avoid the non-linear optimization procedure,
a linear least-squares THC procedure for determining the THC matrices was subsequently
proposed in the literature. \cite{Parrish2012TensorHypercontraction} 
In this case only the $\mathbf{Z}$ matrix is determined by a least-squares fit to the ERI tensor, while the $\mathbf{X}$ matrices are constructed once at the outset and held fixed throughout the least-squares fitting procedure.
Enabled through this seminal contribution, a whole family of least-squares THC (LS-THC)
construction algorithms has been developed, see, e.g.,~Refs.~
\citenum{Hohenstein2012TensorHypercontraction,Parrish2012TensorHypercontraction,Matthews2020ImprovedGrid,Lu2015CompressionElectron,Hu2017InterpolativeSeparable,Dong2018InterpolativeSeparable,Lee2020SystematicallyImprovable,Zhang2025BlockTensor,Hillers-Bendtsen2025LoweringScaling,Hillers-Bendtsen2026NonNegativeLeast}.

\begin{figure*}[htbp]
    \includegraphics[width=\linewidth]{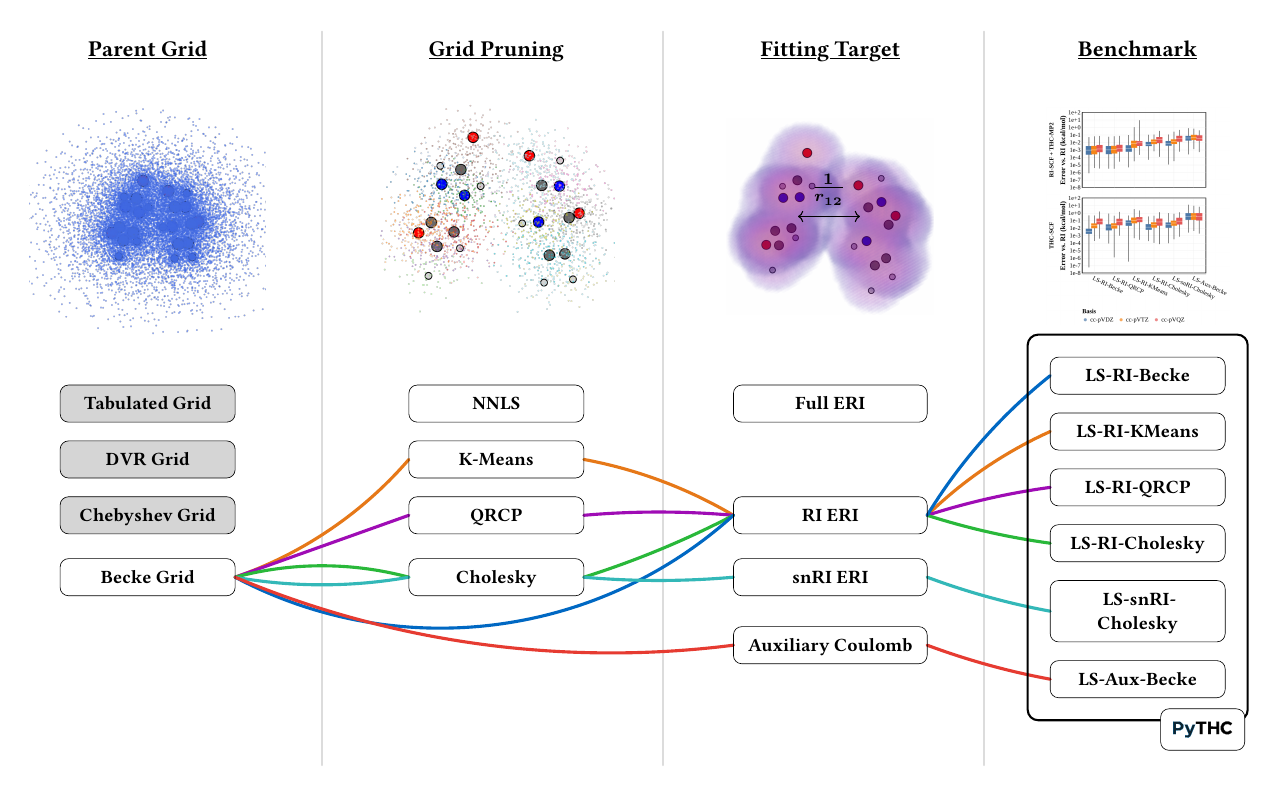}
    \caption{Building blocks of the molecular LS-THC [Eq.~\eqref{eq:thc_decomposition}] construction.
    First, a reference projection space (grid) used for the \(\mathbf{X}\)-matrix  construction is chosen (``Parent Grid''), next this grid can optionally be pruned to reduce the grid-size. Based on this, the \(\mathbf{Z}\)-matrix, for a certain ``Fitting Target'', is determined through a least-squares optimization procedure.
    The different LS-THC variants investigated in this work are indicated by colored lines connecting the various ingredients.
    The naming conventions for the different LS-THC variants benchmarked in this work are summarized in the last columns.}
    \label{fig:thc_family_tree}
\end{figure*}

Overall, the landscape of different LS-THC implementations as well
as their adaptation to electronic structure methods is highly fragmented.
This is reflected in the different naming schemes for these LS-THC
algorithms, which are also denoted as Interpolative Separable Density
Fitting (ISDF),\cite{Lu2015CompressionElectron} RI Real-Space (RI-RS)
\cite{Duchemin2019SeparableResolutionoftheidentity,Duchemin2021CubicScalingAllElectron,Delesma2024BenchmarkingAccuracy}
etc. Throughout this work, we refer
to these linear THC techniques simply as LS-THC. The performance of the various
proposed algorithms in the literature has been assessed on a
wide range of different chemical systems. However, a systematic comparison and
unifying view on the different algorithms is missing. In particular, a common
platform, e.g., software library, that implements the different algorithms in a
unified framework is currently not available. This is the gap we aim to fill
in this work by providing a new open-source
library, called \texttt{PyTHC}, that implements many of the algorithms, which
we used to conduct a differential benchmark (Fig.~\ref{fig:thc_family_tree}).

This work is structured as follows: 
We provide an overview of the different LS-THC methods from a unifying perspective in Sec.~\ref{sec:thc_algos},
details regarding the realization are presented in Sec.~\ref{sec:implementation},
computational details are presented in Sec.~\ref{sec:comp-detail}, and an in-depth comparison of the different procedures is presented in Sec.~\ref{sec:application-validation}.

\section{Theoretical background}
\label{sec:thc_algos}

Throughout this work, occupied MOs are denoted \(i,j,k,l\), virtual MOs are denoted
\(a,b,c,d\), AOs are denoted \(\mu,\nu,\lambda,\sigma\) while \(P,Q,R,S,g\) will be used
for real-space grid points and \(\aux{A},\aux{B},\aux{C},\aux{D}\) for auxiliary basis functions. 

As already mentioned in the introduction, this work compares different
algorithms that construct the THC factorization for the ERI tensor within the linear least-squares (LS) framework. 
This implies that the $\textbf{X}$ matrices are fixed, while $\textbf{Z}$ is determined by 
\begin{align}
    \min_{\mathbf{Z}}\; \left\|\, \eriao - \sum_{P Q} X_{\mu}^P X_{\nu}^P Z^{P Q} X_{\lambda}^Q X_{\sigma}^Q \,\right\|^2.
\end{align}

Ref. \citenum{Parrish2012TensorHypercontraction} showed that linear least
squares fitting can be used to derive an analytical expression for the Coulomb
kernel matrix $\mathbf{Z}$ as
\begin{align}  
    Z^{PS} & = \sum_{\mu \nu \lambda \sigma} \sum_{QR} \left[\mathbf{S}^{- 1}\right]_{P Q} C_{\mu \nu}^Q \eriao C_{\lambda \sigma}^R \left[\mathbf{S}^{- 1}\right]_{R S},
    \label{eq:z_center_matrix}
\end{align}
with
\begin{align}
S_{P Q} & = \sum_{\mu \nu} C_{\mu \nu}^P C_{\mu \nu}^Q \label{eq:SMat}
\\     C_{\mu \nu}^P & = X_{\mu}^P \odot X_{\nu}^P,
\end{align}
where $\mathbf{S}$ will be called the
grid metric-matrix and $\mathbf{C}$ is called the co-density matrix. Here
$\odot$ denotes the face-splitting (Khatri--Rao) product.

To evaluate Eq.~\eqref{eq:z_center_matrix}, a suitable projection
space must be defined to obtain the $\mathbf{X}$ matrix.
A natural choice is a real-space grid. On such a grid, the $\mathbf{X}$ matrix can be computed as
$X_{\mu}^{P} = \sqrt[4]{w_P} \chi_{\mu} (\mathbf{r}_P)$ where $w_P$ denotes the
grid weight at point $P$. 
Since the practical accuracy and efficiency of the LS-THC factorization depends strongly on the grid size, various grid construction and pruning strategies have been investigated in the literature.

A popular choice (for molecular systems) are atom-centered grids, e.g., so-called Becke-type grids, \cite{Becke1988MulticenterNumerical} whose effective grid-size can be further reduced in conjunction with pruning approaches like the
Treutler-Ahlrichs method.\cite{Treutler1995EfficientMolecular}
Alternatively, grids based on Discrete Variable Representation (DVR) have been explored.\cite{Parrish2013DiscreteVariable}
Furthermore, specifically pre-optimized and tabulated atomic grids have been employed. \cite{KokkilaSchumacher2015TensorHypercontraction,Duchemin2019SeparableResolutionoftheidentity,Delesma2024BenchmarkingAccuracy}
We would like to
emphasize that these specially optimized grids require an element- and
basis-set-dependent pre-optimization. 

To avoid rigid pre-optimization of a parent grid, adaptive `on-the-fly' strategies were proposed that reduce the size of a given reference grid.
A common choice for this is to exploit the low-rank structure of $\mathbf{C}$.
For example, Ref.~\citenum{Lu2015CompressionElectron} used the QR decomposition with Column Pivoting (QRCP) for this purpose, which expresses $\mathbf{C}$ as
\begin{align}
	\mathbf{C}^T \mathbf{\Pi} = \mathbf{Q} \mathbf{R}.
\end{align}
Here, $\mathbf{R}$ denotes the upper triangular matrix with descending diagonal elements, 
$\mathbf{Q}$ an orthogonal matrix, and $\mathbf{\Pi}$ is a column pivoting matrix.
The QR decomposition allows for a low-rank approximation, by selecting a set of columns in $\mathbf{R}$ according to a user-defined threshold for their value on the diagonal.
This subset represents the new pruned grid. 
Since standard QRCP on $\mathbf{C} \in \mathbb{R}^{(N_\mathrm{Grid} \times N^2)}$ ($N_\mathrm{Grid}$: Number of grid points; $N$: Size of AO  basis) would result in an $\bigO(N^4)$ complexity step, Ref.\citenum{Lu2015CompressionElectron} employed a matrix compression
scheme to reduce the number of rows from $N^2$ to a number proportional to $N$.
The compression is achieved by applying column-wise FFT and drawing a random
sample (illustrated in SI Alg.~1). However, this scheme (even including the compression) resulted in a large prefactor limiting its applicability.\cite{Dong2018InterpolativeSeparable}

For this reason, Ref.~\citenum{Dong2018InterpolativeSeparable} employed
Centroidal Voronoi Tessellation (CVT) through K-Means clustering using the basis function density as weighting 
(the adaptation of this work is presented in the SI, Alg.~2).
In the K-Means algorithm, cluster centers and cluster-assignment are computed by
iteratively solving the minimization problem
\begin{align}
     \min_{\Gamma_k, \mathbf{c}_k} \;\sum^{K}_k \sum_{\mathbf{r} \in C_k} w(\mathbf{r}) \left\| \mathbf{r} - \mathbf{c}_k \right\|^2,
\end{align}
where $\Gamma$ denotes the $K$ clusters of points $\mathbf{r}$ with $\mathbf{c}$
as their centers and $w$ denotes the weighting function. 
When used on Becke-type grids, Ref.~\citenum{Lee2020SystematicallyImprovable} suggested using the grid's inherent weighting function to simplify subsequent derivations. 
In this procedure, the number of
chosen interpolation points is controlled by the number of initialized cluster centers.
In the method proposed by Ref.~\citenum{Lee2020SystematicallyImprovable}, the K-Means clustering is computed for each atomic partition of the Becke grid, yielding linear scaling with respect to the number of atoms in the system. 
We note that, while this procedure depends on the initial positions of the centroids, we did not observe the large error fluctuations reported in a recent preprint.\cite{Li2026DeterministicTensor}

An alternative grid pruning procedure has been proposed in Ref.~\citenum{Matthews2020ImprovedGrid}.
Realizing that 
\begin{align}
	\mathbf{S} = \mathbf{C} \mathbf{C}^T \approx \mathbf{\Pi} \mathbf{R}^T \mathbf{R} \mathbf{\Pi}^T,
\end{align}
a pivoted Cholesky Decomposition (CD) of the grid metric
\begin{align}
\mathbf{S} = \mathbf{\Pi} \mathbf{L}^T \mathbf{L} \mathbf{\Pi}^T
\label{eq:cholesky}
\end{align}
can be used to select a sparse subgrid. 
Here $\mathbf{L}$ denotes an upper triangular matrix. 
This method was used by many subsequent
publications\cite{Zhao2023OpenShellTensor,Smyser2024UseMultigrids,Pokhilko2024TensorHypercontraction,Yin2025SpatialSignatures,Zhu2026InterpolativeSeparable,Yeh2023LowScalingAlgorithm,Liao2026InterpolativeSeparable,Yang2026InitioManybody}
and was shown to be more efficient than full QRCP while still providing high-fidelity approximations.
The pivoted Cholesky Decomposition only adds an $\bigO(N_{\mathrm{Grid}}^3)$ step to the algorithm in combination with a single parameter to control the fidelity of the
low-rank representation, by stopping after a given self-overlap threshold
on the diagonal of $\mathbf{S}$ defined by (in atomic orbital basis)
\begin{align}
	S_{P P} = \left(\sum_{\mu} \left(X_{\mu}^P\right)^2\right)^2.
\end{align}

Recently, a non-negative least-squares (NNLS) approach was proposed in order to compute custom-tailored grid
weights instead of relying on the standard weights of the Becke grid.\cite{Hillers-Bendtsen2026NonNegativeLeast}

Making use of the pruned or unpruned grids as the projection space, the $\mathbf{Z}$-matrix for the THC decomposition [Eq.~\eqref{eq:thc_decomposition}] is computed through Eq.~\eqref{eq:z_center_matrix}. 
While this is a straightforward linear algebra
computation, the contraction of the co-density with the full ERI (the fitting target)
scales as $\bigO(N^5)$.
To lower this scaling, the RI approximation can be used.\cite{Parrish2012TensorHypercontraction}
Here, $\mathbf{Z}$ is evaluated from
\begin{align}    
 Z^{PS} & = \sum_{\tilde{B}} D_P^{\tilde{B}} D_S^{\tilde{B}},
 \label{eq:ZDDT}
\end{align}
with
\begin{align}
 D_P^{\aux{B}} & = \sum_{Q \aux{A}} \left[\mathbf{S}^{- 1}\right]_{P Q} \left[\mathbf{V}^{-\frac{1}{2}}\right]_{\aux{A} \aux{B}} \sum_{\mu \nu} C_{\mu \nu}^Q \cint{\mu \nu}{\aux{A}}.
 \label{eq:ls_thc_df_fit}
\end{align}
In this case, the contraction of the co-density with the three-center
integral $\cint{\mu \nu}{\aux{A}}$ is the computationally most expensive step, scaling as
$\bigO(N^4)$ (assuming $N_{\mathrm{Aux}} \propto N$).

We note that the $\mathbf{Z}$ matrix can, in principle, be directly evaluated in
$\bigO(N^3)$ complexity through the interpolative decomposition
\begin{align}
C_{\mu \nu} (\mathbf{r}) = \sum_{P'} C_{\mu
\nu}^{P'} \xi_{P'} (\mathbf{r})
\label{eq:isdf_codens}
\end{align}
where $\mathbf{Z}$ is constructed from the interpolation functions $\xi$ 
\begin{align}
Z^{PQ} & = \int \mathrm{d} \mathbf{r}_1 \int \mathrm{d} \mathbf{r}_2 \frac{\xi_P \left(\mathbf{r}_1\right) \xi_Q \left(\mathbf{r}_2\right)}{\left\|\mathbf{r}_1 - \mathbf{r}_2\right\|}.
\end{align}
However, this direct route is of limited applicability for molecular systems \cite{Lee2020SystematicallyImprovable} and will not be further considered in this work. 

From the discussion above, it becomes clear that the total computational cost of the LS-THC factorization depends on the fitting target, the cost of constructing the projection space, and the size of that projection space.
When using the RI approximation as fitting target (in the following denoted as RI-ERI), the overall scaling is still $\bigO(N^4)$ and may therefore dominate the overall cost of the subsequent THC-based electronic structure calculations, e.g., in the case of Hartree--Fock which scales with $\bigO(N^3)$ using a THC-based implementation.\cite{Hu2017InterpolativeSeparable}

To further reduce the scaling, Refs.~\citenum{Zhang2025BlockTensor},\citenum{Zhang2026FormalON3} made use of semi-numerical integration for determining the three-center integrals 
\begin{align}
	\cint{\mu \nu}{\aux{A}} & \approx \sum_g \chi_\mu (\mathbf{r}_g) \chi_\nu (\mathbf{r}_g) \int \frac{\chi_{\aux{A}} (\mathbf{r})}{\| \mathbf{r} - \mathbf{r}_g \|} \mathrm{d} \mathbf{r},
	\label{eq:sri-approx}
\end{align}
where the integration grid is indexed by $g$.
Based on this, Eq.~\eqref{eq:ls_thc_df_fit} can be reformulated
\begin{align}
	D_P^{\aux{B}} & = \sum_{Q \aux{A}} \left[\mathbf{S}^{- 1}\right]_{P Q} \left[\mathbf{V}^{-\frac{1}{2}}\right]_{\aux{A} \aux{B}} \sum_{\mu \nu g} C_{\mu \nu}^Q C_{\mu \nu}^g \int \frac{\chi_{\aux{A}} (\mathbf{r})}{\| \mathbf{r} - \mathbf{r}_g \|} \mathrm{d} \mathbf{r} \nonumber \\
	& = \sum_{Q \aux{A}} \left[\mathbf{S}^{- 1}\right]_{P Q} \left[\mathbf{V}^{-\frac{1}{2}}\right]_{\aux{A} \aux{B}} \sum_{g} S_g^Q \int \frac{\chi_{\aux{A}} (\mathbf{r})}{\| \mathbf{r} - \mathbf{r}_g \|} \mathrm{d} \mathbf{r},
\end{align}
with
\begin{align}
S_g^Q = \sum_{\mu \nu} C_{\mu \nu}^Q C_{\mu \nu}^g = \sum_{\mu \nu} \left(X_{\mu}^Q X_{\mu}^g\right) \left(X_{\nu}^Q X_{\nu}^g\right),
\end{align}
which only involves $\bigO(N^3)$ contractions
and therefore reduces the overall scaling. In the following, we will refer to
this fitting target as semi-numerical RI (snRI).

Similarly, the complexity can be reduced to $\bigO(N^3)$ by replacing the full ERI with the Coulomb metric $\mathbf{V}$ as a
fitting target.\cite{Hillers-Bendtsen2025LoweringScaling,Hillers-Bendtsen2025AcceleratingHartree} 
In this case, the Coulomb half-kernel from Eq.~\eqref{eq:ZDDT} is determined as
\begin{align}    
 D_P^{\aux{B}} & = \sum_{Q \aux{A}} \left[\mathbf{\bar{S}}^{- 1}\right]_{P Q} X_{\aux{A}}^Q \left[\mathbf{V}^{-\frac{1}{2}}\right]_{\aux{A} \aux{B}}
 \label{eq:ls_thc_aux_fit}
\end{align}
with
\begin{align}
 \bar{S}_{P Q} & = \sum_{\aux{A}} X_{\aux{A}}^P X_{\aux{A}}^Q.
 \label{eq:SMat2}
\end{align}

On top of the above considerations, sparsity of the AO can be additionally exploited.\cite{Song2016AtomicOrbitalbased,Duchemin2019SeparableResolutionoftheidentity,Wang2023SparsityElectron,Lee2020SystematicallyImprovable,Zhang2025BlockTensor} 
\section{Implementation details}
\label{sec:implementation}

From Sec.~\ref{sec:thc_algos}, it becomes clear that the LS-THC factorization
in molecular systems consists of three main steps: (i) the choice of a
projection space (`parent grid'), (ii) optional grid pruning, and (iii) fitting
to the chosen target for the construction of $\mathbf{Z}$.
Possible options for these three steps are summarized in Fig.~\ref{fig:thc_family_tree}. 
The white boxes in Fig.~\ref{fig:thc_family_tree} have been implemented as part of this work.
The resulting open-source library, \texttt{PyTHC}, allows for modular construction of the various LS-THC algorithms. 
The algorithms investigated here are indicated by lines in different colors connecting the different ingredients.
To differentiate the algorithms throughout this work, we use the following naming convention:
LS-\textit{FittingTarget}-\textit{PruningStrategy}.
Note that additional sparsity considerations are not investigated in this work.
Furthermore, we did not include the recent NNLS approach\cite{Hillers-Bendtsen2026NonNegativeLeast} in our benchmark, although an unoptimised preliminary implementation is available in \texttt{PyTHC} (see Fig.~\ref{fig:thc_family_tree}).

The basic structure of all procedures for the three fitting targets (RI-ERI, Aux-Coulomb, snRI-ERI) is outlined in Algs.~\ref{algo-fit-df-eri},\ref{algo-fit-aux-coulomb}
and~\ref{algo-fit-ls-thc-btd}.
In all cases, the first step is to build a reference grid (e.g., a Becke grid) and evaluate the basis functions on this grid.
For this, we rely on the \texttt{PySCF} implementations.\cite{Sun2018PySCFPythonbased,Sun2020RecentDevelopments}
The resulting $\mathbf{X}$-matrix is then optionally pruned using one of the three pruning strategies.
Detailed pseudocode for these pruning strategies is provided in the SI Sec.~I.

We also implemented unrestricted versions of the THC building algorithms.
In the case of fitting the RI or snRI-ERI, different $\mathbf{X}$ matrices
exist for the $\alpha$- and $\beta$-spin cases leading to different Coulomb
matrices, namely $\mathbf{Z}_{\alpha \alpha}$ $\mathbf{Z}_{\beta \beta}$ and
$\mathbf{Z}_{\alpha \beta}$. As QRCP based pruning operates in the AO basis,
the grid pruning remains unchanged in this case. For Cholesky-based pruning, an
additional adaptation can be made to only fit the occupied-virtual (ov) block,
by decomposing the metric matrix defined by $S^{P Q} = \sum_{i a} X_i^P X_a^P
X_i^Q X_a^Q$ instead of the AO metric. For K-Means based pruning the weighting
functions uses  the MO transformed $\mathbf{X}$ matrix. This can also be adopted for the
unrestricted case, yielding different grids for the two spin cases.

\begin{algfigure}
\renewcommand{\figurename}{Algorithm} % Applies to all captions in this float
\caption{LS-THC with RI-ERI fitting target}
\label{algo-fit-df-eri}
\begin{algorithmic}[1] 
\Procedure{LS-RI-[*]}{}
    \State $\mathbf{g}, \mathbf{w} \gets \text{build Becke grid and weights}$
    \Statex
    \State $\text{// Prune grid with Cholesky, K-Means or QRCP (Optional)}$
    \State $\mathbf{g'} \gets \text{prune } \mathbf{g} \text{ with error threshold } \epsilon$
    \Statex
    \State $X^P_\mu \gets \sqrt[4]{w_P} \cdot \chi_\mu(g'_P)$ 
    \State $\mathbf{C} \gets \mathbf{X} \odot \mathbf{X}$
    \State $G_{\aux{A}\aux{B}} \gets (\aux{A}|\aux{B})$
    \State $V_{\mu \nu}^{\aux{A}} \gets (\mu \nu|\aux{A})$
    \State $\mathbf{V}^{-\frac{1}{2}} \gets \operatorname{Cholesky}(\mathbf{V})$
    \State $K_{\aux{A}}^{P} \gets \sum_{\mu \nu} V^{\aux{A}}_{\mu \nu} C_{\mu \nu}^P$
    \State $Y_{\aux{A}}^{P} \gets \sum_{\aux{B}} K_{\aux{B}}^P  [V^{-\frac{1}{2}}]_{\aux{A} \aux{B}}$
    \Statex
    \State $S^{PQ} \gets \sum_{\mu \nu} C_{\mu\nu}^P C_{\mu\nu}^Q$
    \State $\mathbf{S}^+ \gets \operatorname{PseudoInvert}(\mathbf{S})$
    \State $\mathbf{D} \gets \mathbf{S}^+ \mathbf{Y}$
    \State $\mathbf{Z} \gets \mathbf{D}^T \mathbf{D}$
    \Statex
    \State \Return $\mathbf{X}, \mathbf{Z}$
\EndProcedure
\end{algorithmic}

\caption{LS-THC with snRI fitting target}
\label{algo-fit-ls-thc-btd}
\begin{algorithmic}[1] 
\Procedure{LS-snRI-[*]}{}
    \State $\mathbf{g}, \mathbf{w} \gets \text{build Becke grid and weights}$
    \State $\mathbf{g}^\mathrm{RI}, \mathbf{w}^\mathrm{RI} \gets \text{build Becke grid for} \cint{\mu \nu}{\aux{A}} \text{ fit}$
    \Statex
    \State $\text{// Prune grid with Cholesky, K-Means or QRCP (Optional)}$
    \State $\mathbf{g'} \gets \text{prune } \mathbf{g} \text{ with error threshold } \epsilon$
    \State $X^P_\mu \gets \sqrt[4]{w_P} \cdot \chi_\mu(g'_P)$ 
    \State $X'^P_\mu \gets \sqrt[4]{w_P^\mathrm{RI}} \cdot \chi_\mu(g^\mathrm{RI}_P)$
    \Statex
    \State $\mathbf{C} \gets \mathbf{X} \odot \mathbf{X}$
    \State $\mathbf{C'} \gets \mathbf{X'} \odot \mathbf{X'}$
    \State $S^{PQ} \gets \sum_{\mu \nu} C_{\mu\nu}^P C_{\mu\nu}^Q$
    \State $\mathbf{S}^+ \gets \operatorname{PseudoInvert}(\mathbf{S})$
    \Statex
    \State $V_{\aux{A}\aux{B}} \gets (\aux{A}|\aux{B})$
    \State $\mathbf{V}^{-\frac{1}{2}} \gets \operatorname{Cholesky}(\mathbf{V})$
    \Statex
    \State $V^{\aux{A}}_g \gets \sum_{\aux{B}} [\mathbf{V}^{-\frac{1}{2}}]_{\aux{A} \aux{B}} \left(\sqrt{w^\mathrm{RI}_g} \int \mathrm{d} \mathbf{r}\; \chi_{\aux{B}}(\mathbf{r})/\| \mathbf{r} - \mathbf{g}^\mathrm{RI}_g \| \right)$
    \State $S_g^Q \gets \sum_{\mu \nu} [\mathbf{C'}]_{\mu \nu}^g C_{\mu \nu}^Q$
    \State $B_Q^{\aux{A}} \gets \sum_{g} V_g^{\aux{A}} S_g^Q$
    \Statex
    \State $D^{\aux{A}}_P \gets \sum_{Q} [\mathbf{S}^+]^{PQ} B_Q^{\aux{A}}$
    \State $\mathbf{Z} \gets \mathbf{D}^T \mathbf{D}$
    \Statex
    \State \Return $\mathbf{X}, \mathbf{Z}$
\EndProcedure
\end{algorithmic}
\end{algfigure}

\begin{algfigure}
\caption{LS-THC with Auxiliary-Coulomb fitting target}
\label{algo-fit-aux-coulomb}
\begin{algorithmic}[1] 
\Procedure{LS-Aux-[*]}{}
    \State $\mathbf{g}, \mathbf{w} \gets \text{build Becke grid and weights}$
    \Statex
    \State $\bar{X}_{\aux{A}}^P \gets \sqrt{w_P} \cdot \chi_{\aux{A}}(g'_P)$
    \State $V_{\aux{A}\aux{B}} \gets (\aux{A}|\aux{B})$
    \State $\mathbf{L} \gets \text{Cholesky}(\mathbf{V})$
    \Statex
    \State $\mathbf{\bar{S}} \gets \bar{\mathbf{X}}^T \mathbf{\bar{X}} + \lambda\mathbb{I}$
    \State $\mathbf{M} \gets \operatorname{solve}(\mathbf{\bar{S}}, \mathbf{L})$
    \State $\mathbf{D}^T \gets \mathbf{\bar{X}}\mathbf{M}$
    \State $\mathbf{Z} \gets \mathbf{D}^T \mathbf{D}$
    \Statex
    \State $X^P_\mu \gets \sqrt[4]{w_P} \cdot \chi_\mu(g'_P)$ 
    \State \Return $\mathbf{X}, \mathbf{Z}$
\EndProcedure
\end{algorithmic}
\end{algfigure}

Additional specific implementation choices for the different THC algorithms as part of this work are summarized in the following:
\begin{itemize}
    \item \textbf{LS-[*]-Cholesky}: Instead of the standard greedy Cholesky Decomposition used in Ref.~\citenum{Matthews2020ImprovedGrid}, we employed a version with randomized block pivoting based on Ref.~\citenum{Epperly2025EmbraceRejection} (Accelerated-RPCholesky). It allows lazy evaluation of $\mathbf{S}$ thus lowering the memory requirement of the grid pruning step from $\bigO(N_{\mathrm{Grid}}^2)$ to $\bigO(N_{\mathrm{Grid}} \cdot r)$, $r$ being the approximate numerical rank of the $\mathbf{S}$ matrix. As suggested in  Ref.~\citenum{Matthews2020ImprovedGrid}, we compute the Coulomb half-kernel from Eq.~\eqref{eq:ls_thc_df_fit}
    \begin{align}
       \mathbf{D} = \mathbf{S}^{-1} \mathbf{Y}
    \end{align} using the Cholesky factors $\mathbf{L}$ [Eq.~\eqref{eq:cholesky}] for
    \begin{align}
        \mathbf{L}^T \mathbf{A} &= \mathbf{Y}\; \text{solve for } \mathbf{A}  
        \label{eq:cd1} \\
        \mathbf{L} \mathbf{D} &= \mathbf{A}\; \text{solve for } \mathbf{D}
        \label{eq:cd2}
    \end{align}
    thus avoiding the direct pseudoinversion of the metric matrix $\mathbf{S}$. Here, $\mathbf{Y}$ represents the contraction of the co-density matrix, fitting target and the separable Coulomb metric.

    \item \textbf{LS-[*]-QRCP}: Ref.~\citenum{Lu2015CompressionElectron} introduced two separate error control parameters: $k_\mathrm{Rows}$ determined how many rows of the Fourier transformed matrix are kept in multiples of $N$ (as the number of atomic orbitals) and $\epsilon$ controlled how many pivots of the QRCP are chosen as interpolation points. We found this approach complicated the parameterization and therefore derived the $k_\mathrm{Rows}$ parameter from the $\epsilon$ parameter using
    \begin{align}
        k_\mathrm{Rows} = \max\!\left(1, \left\lceil -\frac{10}{3} \log_{10}{\epsilon} \right\rceil \right).
        \label{eq:qrcp_krows}
    \end{align} 
    Setting $\epsilon=10^{-6}$ leads to $k_\mathrm{Rows}=20$ as used in Ref.~\citenum{Lu2015CompressionElectron}.

    \item \textbf{LS-[*]-KMeans}: We used the following weighting function in the clustering procedure
    \begin{align}
    w\left(\mathbf{r}_P\right) = \sum_{\mu} \left(\;X_{\mu}^P\;\right)^2.
    \label{eq:weighting_function}
    \end{align}
    Furthermore, the number of allocated interpolation points was dynamically scaled based on the proportion of the grid points per atom relative to the total grid size. To efficiently map the resulting cluster centers back to their nearest grid neighbors, a $k$-d-Tree lookup structure was implemented, providing $\mathcal{O}(\log N_\mathrm{Grid})$ query complexity with a modest memory footprint of $\mathcal{O}(3 \cdot N_\mathrm{Grid})$.

    \item \textbf{LS-Aux-[*]}: Instead of using the LQ factorization, we employed Ridge regularization\cite{Hoerl1970RidgeRegression} with a damping parameter of $\lambda = 10^{-8}$ to solve the system
    \begin{align}
      \mathbf{D}^T = \mathbf{\bar{X}}(\mathbf{\bar{X}}^T\mathbf{\bar{X}} + \lambda\mathbb{I})^{-1}\mathbf{L}  
    \end{align} where
    \begin{align}
    \bar{X}_{\aux{A}}^P = \sqrt{w_P} \cdot \chi_{\aux{A}}(g'_P)
    \end{align} is the auxiliary collocation matrix (see Alg.~\ref{algo-fit-aux-coulomb}). This is equivalent to the minimum-norm solution provided by the LQ factorization (where $\lambda=0$) but it provides higher numerical stability if $\mathbf{\bar{X}}$ is ill-conditioned, as was the case in some systems in which we used Even Tempered Basis sets (ETBs).
    \end{itemize}

To evaluate the accuracy of the THC decomposition and the error incurred
by the fitting, we implemented THC based Hartree-Fock (HF) Fock matrix
construction and second-order M{\o}ller-Plesset perturbation theory (MP2).
The Coulomb $\mathbf{J}$ and exchange $\mathbf{K}$ matrices entering the HF Fock matrix are computed as
\begin{align}
\label{eq:thc-scf-j}
J_{\lambda \sigma} =  \sum_{\mu \nu} \sum_{P Q} \rho_{\mu \nu} X_\mu^P X_\nu^P Z^{P Q} X_\lambda^Q X_\sigma^Q \\
\label{eq:thc-scf-k}
K_{\lambda \sigma} =  \sum_{\mu \nu} \sum_{P Q} \rho_{\mu \nu} X_\mu^P X_\lambda^P Z^{P Q} X_\nu^Q X_\sigma^Q,
\end{align}
where $\rho_{\mu \nu}$ denotes the one-particle reduced density matrix. 

To calculate the MP2 correlation energy, we make use of the
Laplace transformation technique.\cite{Haser1992LaplaceTransform} 
The spin-summed MP2 energy reads
\begin{align}
    E_{\text{MP2}} & = - \sum_{ijab} \frac{(ia|jb) \left[ 2(ia|jb) - (ib|ja) \right]}{\epsilon_a + \epsilon_b - \epsilon_i - \epsilon_j} \\ 
    & = - \sum_{i j a b}\left[ 2 \left(i a | j b\right)^2 - \left(i a | j b\right)\left(i b | j a\right)\right]\Delta_{i j}^{a b} \\
    & = -2 \left[\sum_{i j a b} \left(i a | j b\right)^2 \Delta_{i j}^{a b}\right] + \left[\sum_{i j a b} \left(i a | j b\right)\left(i b | j a\right) \Delta_{i j}^{a b}\right] \\
    & = -2 \cdot J^{\mathrm{MP2}} + K^{\mathrm{MP2}},
\end{align}
where the inverse denominator $\mathbf{\Delta}$ is approximated by numerical integration, employing the Laplace transformation
\begin{align}
    \Delta_{i j}^{a b} = \int_0^{\infty} \mathrm{d} t\; e^{\epsilon_i t} e^{\epsilon_j t} e^{-\epsilon_a t} e^{-\epsilon_b t} \approx \sum_v^{N_{\mathrm{Lapl}}} w_v \tau_v^i \tau_v^j \tau_v^a \tau_v^b.
\end{align}

Combined with the THC approximation of the ERI, we can write the MP2 energy as \cite{Hohenstein2012TensorHypercontraction}
\begin{align}
    J^{\mathrm{MP2}} & = \sum_v \sum_{i j a b} \sum_{PQRS} \tau_v^i \tau_v^j \tau_v^a \tau_v^b~X_i^P X_a^P Z^{P Q} X_j^Q X_b^Q~X_i^R X_a^R Z^{R S} X_j^S X_b^S \label{eq:JMP2}\\
    K^{\mathrm{MP2}} & = \sum_v \sum_{i j a b} \sum_{PQRS} \tau_v^i \tau_v^j \tau_v^a \tau_v^b~X_i^P X_a^P Z^{P Q} X_j^Q X_b^Q~X_i^R X_b^R Z^{R S} X_j^S X_a^S \label{eq:KMP2}
\end{align}
where the integration weights $w_v$ are contracted into the $\mathbf{\tau}$ matrices.
This results in a formal scaling of $\bigO(N^4)$ for the MP2 energy evaluation.
If one chooses to neglect $ K^{\mathrm{MP2}}$, as it is done in Scaled Opposite-Spin MP2
(SOS-MP2),\cite{Jung2004ScaledOppositespin} the overall complexity is further lowered to $\bigO(N^3)$.

We employ the tensor network contraction library
\texttt{cotengra}\cite{Gray2021HyperoptimizedTensor} for both the HF $\mathbf{J}$ and $\mathbf{K}$
[Eqs.\eqref{eq:thc-scf-j}-\eqref{eq:thc-scf-k}] contractions. As the dimensions
of $\rho_{\mu\nu}$ do not change through iterations, we can perform the
optimization once at the beginning of the algorithm and reuse it in every
iteration. For the MP2 calculations, we implemented a version of the contraction
path suggested in Ref.~\citenum{Hohenstein2012TensorHypercontraction}.

Most operations involved in the given algorithms are implemented
using standard NumPy.\cite{Harris2020ArrayProgramming} 
Tensor contractions with more than two input tensors are handled
by the \texttt{opt-einsum}\cite{Smith2018Opt_einsumPython} library.
More advanced algorithms like FFT, k-d-Tree and QRCP are sourced from
SciPy\cite{Virtanen2020SciPy10} while we used the K-Means Clustering algorithm
from scikit-learn.\cite{Pedregosa2011ScikitlearnMachine}

\section{Computational Details}
\label{sec:comp-detail}

All THC-MP2 calculations were based on a RI mean-field calculation in order  to
properly isolate the error introduced by THC.  We used 10 grid points in
the numerical quadrature of the Laplace transformation. The grid was built using
\texttt{PySCF}'s routines for modified Gauss-Legendre quadrature. 
All LS-RI-[*] methods use the ov block of the three center RI tensor (i.e., $\cint{ia}{\aux{A}}$) for
the MP2 calculations and the full AO integrals (i.e., $\cint{\mu \nu}{\aux{A}}$)
for the HF-SCF algorithm.

The SCF was converged using the \texttt{PySCF} default settings with the convergence
threshold set to $10^{-9}$ Hartree throughout.
We employed a SAD initial guess and the maximum number of SCF cycles was set to 200.
In THC-based HF-SCF calculations, the first SCF cycles are performed
using the RI-ERI until the difference in energy is below 0.1 Hartree. 
THC is used for the subsequent convergence of the SCF.
For all THC methods the reference real-space grid used the level 0 setting of the \texttt{PySCF} DFT grid builder.
Furthermore, Treutler--Ahlrichs pruning\cite{Treutler1995EfficientMolecular} was used throughout.
We refer to this parent grid simply as ``Becke grid'' in the following.
In LS-snRI-[*] we used level 1 grids to approximate the three-center RI integrals [Eq.~\eqref{eq:sri-approx}].

All tests were conducted in correlation consistent basis sets cc-pVDZ,
cc-pVTZ and cc-pVQZ.\cite{Dunning1989GaussianBasis}
We used the corresponding RI fitting basis sets
cc-pVXZ-RI\cite{Weigend2002EfficientUse} in all steps that involve the
RI-ERI. For systems including elements heavier than (and including)
antimony we used the heavy element versions of the corresponding basis
sets\cite{Papajak2009EfficientDiffuse} (i.e., cc-pVDZ-PP for the cc-pVDZ
basis) as well as Effective Core Potentials (ECPs). If elements are missing
from the auxiliary basis sets, we used \texttt{PySCF} to derive an Even Tempered
Basis (ETB)\cite{Bardo1973EventemperedAtomic} with the parameter
$\beta=2$.

For all pseudoinversion procedures we used the eigen-decomposition based
algorithm suggested in Ref.~\citenum{Parrish2012TensorHypercontraction} with
the threshold set to $10^{-10}$. All computations were executed on 2 AMD EPYC 9654 CPUs (192 cores in total) utilizing 768 gigabytes of RAM.

\section{Application and validation}
\label{sec:application-validation}

\begin{table*}[tbph]
\centering
\caption{Default values of the error control parameters for all LS-THC methods for a given parent grid. The parameters for the AO and MO (ov block only) ERI fit were chosen as they minimized errors in the respective HF-SCF and MP2 calculations. These parameters are used throughout this work.}
\label{tab:parameters_desc}
\begin{tabular}{l l p{7cm} c c}
\toprule
\midrule
\textbf{Method} & \textbf{Parameter} & \textbf{Description} & \textbf{Value MO} & \textbf{Value AO} \\ \midrule
LS-[*]-Cholesky & Cholesky Threshold & Minimum size of diagonal element during CD pivot & $10^{-5}$  & $10^{-8}$\\ \midrule
LS-[*]-Kmeans & IPs per $N_{aux}$ & Number of cluster centers used in K-Means as multiple of the auxiliary basis set size & $3.5$ & $5.5$ \\ \midrule
LS-[*]-QRCP & Tolerance & Minimum size grid point value on diagonal of $\mathbf{R}$ + Kept rows in matrix compression [Eq.~\eqref{eq:qrcp_krows}] & $10^{-4}$ & $10^{-6}$\\ \midrule
LS-Aux-[*] & Auxiliary Basis & Basis set used as fitting target & cc-pV5Z & cc-pV5Z\\
& Regularization & Ridge regularization parameter when finding the minimum-norm solution & $10^{-4}$ & $10^{-8}$ \\ 
\midrule
\bottomrule
\end{tabular}
\end{table*}

\begin{figure*}[tbph]
    \includegraphics[width=\linewidth]{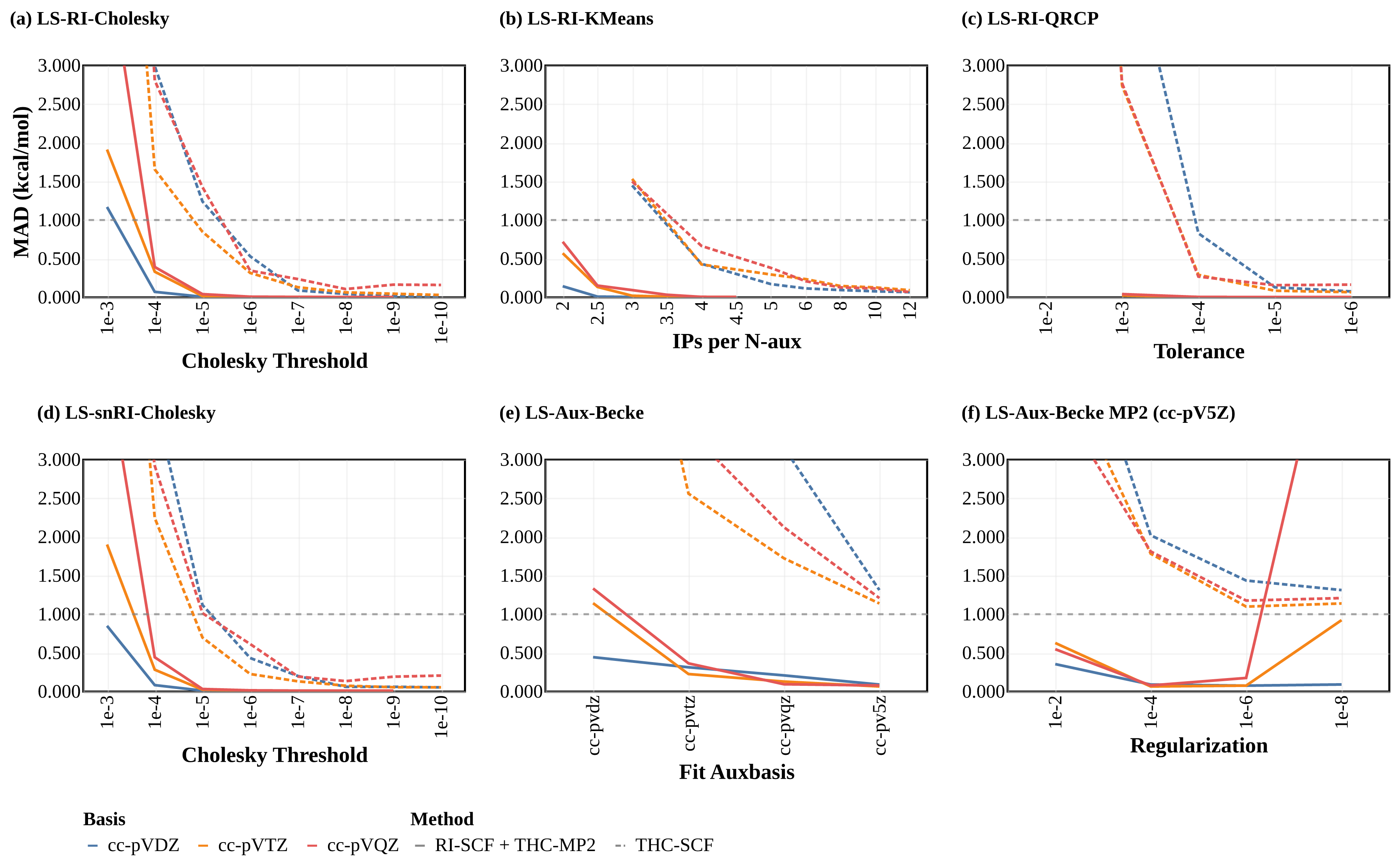}
    \caption{Mean Absolute Deviation (MAD) of the THC-SCF and RI-SCF + THC-MP2 reaction energy barriers computed on the Diet-GMTKN55 30-reaction subset in the cc-pVDZ, cc-pVTZ and cc-pVQZ basis sets with differing error control parameters. Ordering from left to right implies a higher fidelity setting. For LS-RI-[Cholesky/Kmeans/QRCP] this is achieved by a larger pruned grid relative to parent grid, while for LS-Aux-Becke a bigger auxiliary fitting basis is chosen. 
    The 1 kcal/mol threshold is indicated as dotted gray line.}
    \label{fig:parameter_search}
\end{figure*}

We aim to thoroughly evaluate the performance of the
different LS-THC algorithms investigated in this work: 
LS-RI-Becke, LS-RI-KMeans, LS-RI-QRCP, LS-RI-Cholesky, LS-snRI-Cholesky, and LS-Aux-Becke.
For this, we start with a parameter search for the grid-pruning procedures (Sec.~\ref{sec:parameter-search}), next the different methods are benchmarked (Sec.~\ref{sec:benchmarking}), and finally their computational performance is evaluated (Sec.~\ref{sec:timings}).

For the parameter search and benchmarking, we used a curated subset of the 
GMTKN55\cite{Goerigk2017LookDensity} main-group element reaction dataset called
Diet-GMTKN55.\cite{Gould2018DietGMTKN55} 
GMTKN55 offers an extensive testing ground for molecular electronic structure calculations. 

Diet-GMTKN55 assembles subsets of 30 to
150 energy differences by employing a stochastic genetic approach.\cite{Gould2018DietGMTKN55}
These smaller sets reproduce the main trends of the full GMTKN55 dataset at a fraction of the computational cost.
In our analysis, we calculated the energy differences for the datasets based on the THC-based methods and compared them to the reference values obtained from electronic structure calculations employing the RI approximation.

\subsection{Parameter search}
\label{sec:parameter-search}

All of our implemented algorithms, except LS-RI-Becke, expose
one or more parameters for the given parent grid.
A description of the different parameters for the THC procedures is provided in
Tab.~\ref{tab:parameters_desc}. 
We used the 30 energy difference subset of Diet-GMTKN55 to sample the parameter space of our THC procedures for both the HF-SCF and MP2 calculations. 
The corresponding mean absolute deviation (MAD) of the benchmark set is shown in Fig.~\ref{fig:parameter_search} for the different parameters and basis sets.

For all grid pruning methods (Cholesky, K-Means and QRCP), Fig.~\ref{fig:parameter_search} a), b), c) and d) show that the MAD in all basis sets converges towards the RI reference values for both the THC-based HF-SCF and MP2 calculations when the grid size is increased (handled through the individual parameters). We note that similar convergence behavior has also been observed in the literature for the LS-RI-KMeans and LS-RI-Cholesky methods, albeit for less comprehensive benchmark sets.\cite{Matthews2020ImprovedGrid,Lee2020SystematicallyImprovable}

In the case of LS-Aux-Becke, Fig.~\ref{fig:parameter_search} e) and f), the central parameter is the size of the auxiliary fitting basis used for the Coulomb metric, which represents an approximation to the ERI tensor.
We find that the MAD converges when increasing the auxiliary basis size from cc-pVDZ-RI to cc-pV5Z-RI for the HF-SCF and MP2 calculations.
However, the MAD remains above 1 kcal/mol in the case of the HF-SCF calculations for all basis sets.
Furthermore, we observed larger errors in systems with ETB in the auxiliary fitting basis.
For this reason, we investigated the role of the regularization parameter for the 30-reaction subset [Fig.~\ref{fig:parameter_search} f)].
It becomes clear that the MP2 calculations require a stronger regularization, whereas the HF-SCF calculations show increased MADs with stronger regularization.
We therefore suggest a regularization of $10^{-4}$ for MP2 and $10^{-8}$ for the HF-SCF calculations.

The chosen set of default parameters for all LS-THC methods is summarized in Tab.~\ref{tab:parameters_desc} and will be used in the remainder of this work.

\subsection{Benchmarking}
\label{sec:benchmarking}

\begin{figure}[tbp]
    \includegraphics[width=.9\linewidth]{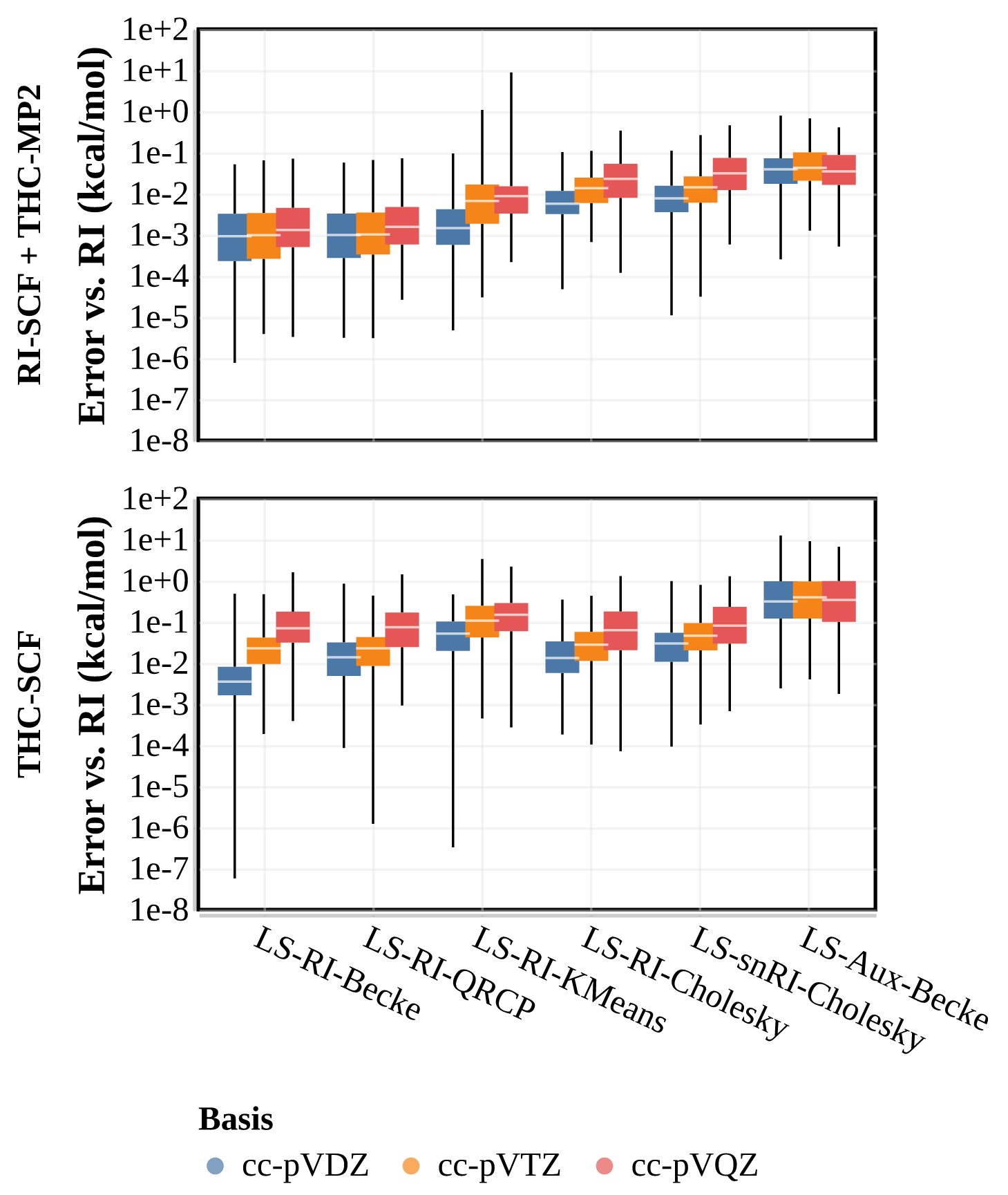}
    \caption{Box-plot distribution of the absolute errors (relative to RI reference values) produced by each LS-THC construction method across the three basis sets: cc-pVDZ, cc-pVTZ, and cc-pVQZ, evaluated on the 150-reaction Diet-GMTKN55 subset.}
    \label{fig:relative_errors_diet_gmtkn55}
\end{figure}

\begin{figure}[tbp]
    \includegraphics[width=.9\linewidth]{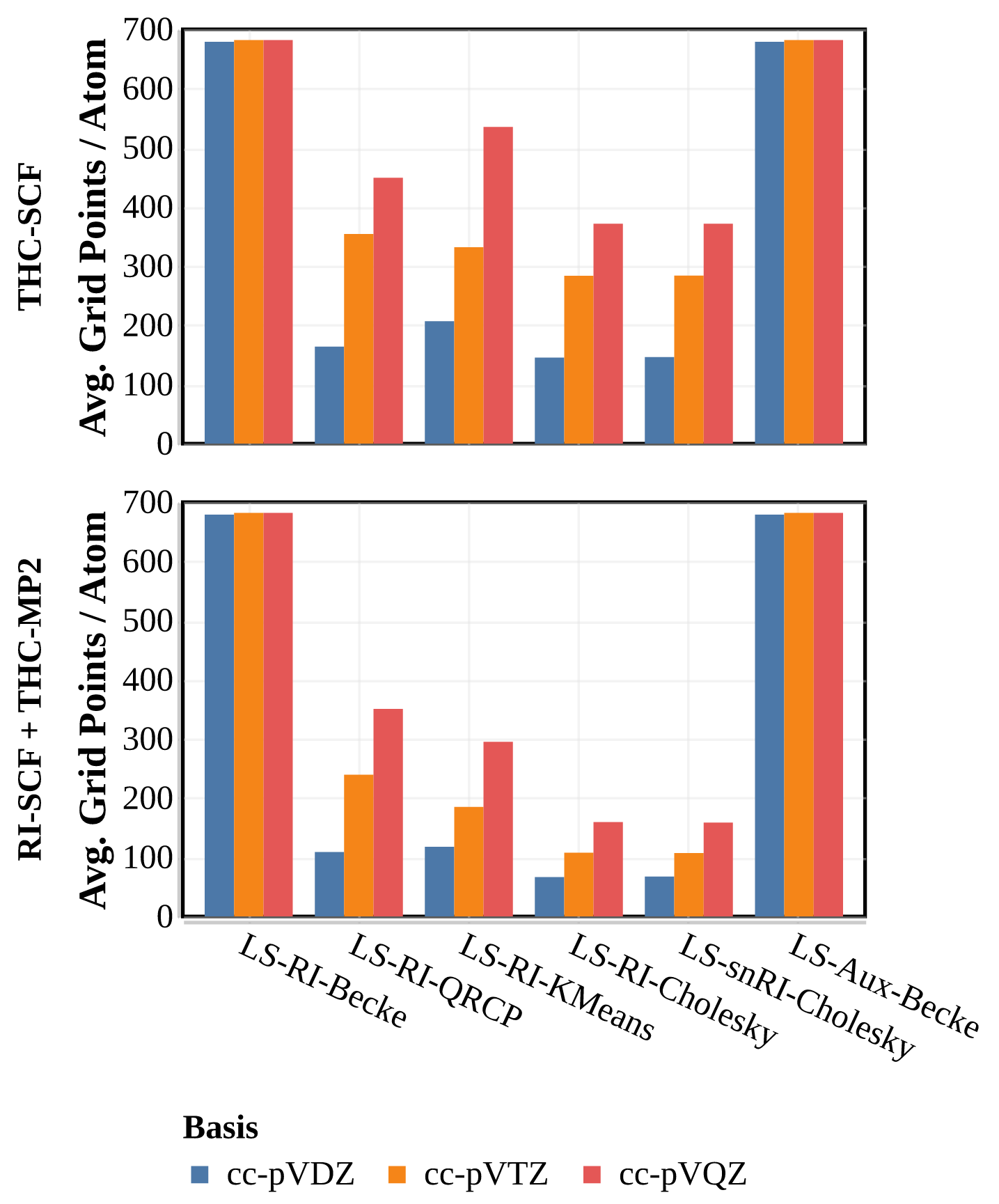}
    \caption{Average number of THC quadrature grid points per atom achieved by each LS-THC construction method on the 150-reaction Diet-GMTKN55 subset. All LS-[*]-Becke methods employ the full molecular Becke-type grid. THC-SCF grids were constructed to approximate the atomic orbital electron-repulsion integral, whereas THC-MP2 grids target only the occupied--virtual (ov) block in molecular orbital basis. Grid points and atom counts are aggregated over all reactants and products of each reaction.}
    \label{fig:grid_sizes_per_atom}
\end{figure}

\begin{table}
\renewcommand{\arraystretch}{1.2}
\caption{Summary statistics Mean Absolute Deviation (MAD), Median Absolute Deviation (MED), Interquartile Range (IQR) and Maximum Absolute Deviation (MAX) of the implemented LS-THC methods on the 150-reaction subset of Diet-GMTKN55 for the cc-pVQZ basis in kcal/mol.}
\label{tab:energy_errors}
\begin{tabular}{llrrrr}
\toprule
 &  & MAD & MED & IQR & MAX \\
Type & Method &  &  &  &  \\
\midrule
\multirow[t]{6}{*}{THC-SCF} & LS-RI-Becke & 0.165 & 0.069 & 0.145 & 1.591 \\
 & LS-RI-QRCP & 0.154 & 0.074 & 0.143 & 1.418 \\
 & LS-RI-KMeans & 0.246 & 0.147 & 0.225 & 2.198 \\
 & LS-RI-Cholesky & 0.147 & 0.062 & 0.157 & 1.281 \\
 & LS-snRI-Cholesky & 0.174 & 0.080 & 0.200 & 1.275 \\
 & LS-Aux-Becke & 0.751 & 0.340 & 0.876 & 6.673 \\
\cline{1-6}
\multirow[t]{6}{*}{RI-SCF + THC-MP2} & LS-RI-Becke & 0.004 & 0.001 & 0.004 & 0.070 \\
 & LS-RI-QRCP & 0.004 & 0.002 & 0.004 & 0.072 \\
 & LS-RI-KMeans & 0.179 & 0.009 & 0.012 & 8.853 \\
 & LS-RI-Cholesky & 0.041 & 0.023 & 0.045 & 0.338 \\
 & LS-snRI-Cholesky & 0.056 & 0.031 & 0.061 & 0.459 \\
 & LS-Aux-Becke & 0.066 & 0.035 & 0.070 & 0.405 \\
\cline{1-6}
\bottomrule
\end{tabular}
\end{table}

Building on the parameters established in the preceding section, we extend the benchmark evaluation to the full 150-reaction Diet-GMTKN55 subset.  The resulting error distributions are presented in Fig.~\ref{fig:relative_errors_diet_gmtkn55}, the corresponding average grid sizes per atom in Fig.~\ref{fig:grid_sizes_per_atom}, and summary statistics for the cc-pVQZ basis in Tab.~\ref{tab:energy_errors} (a complete basis-set breakdown is provided in SI, Tab.~I).
We note that for two systems [G21EA.EA\_23n (neutral phosphorus monoxide) and RC21.7e (a cationic hydrocarbon) in the cc-pVDZ basis] neither the RI-SCF reference nor any of the THC-based SCF variants converged within the prescribed limit of 200 iterations.
These two systems are excluded from all subsequent analyses.
Furthermore, the datasets UPU23, ISOL24, C60ISO and INV24 included systems with over 3000 basis functions in cc-pVQZ and could not be completed with all methods on the given hardware (most critically with the full LS-RI-Becke method). 
We therefore excluded them from the comparative analysis to avoid bias in the reported statistics. 
We included all energy differences that could be computed in an overview figure in the SI (Sec.~II, Fig.~4).

The THC-MP2 results from Tab.~\ref{tab:energy_errors} and Fig.~\ref{fig:relative_errors_diet_gmtkn55} reveal the following trends: 
All methods fitting either the full RI-ERI or the semi-numerical RI (snRI) approximation kept errors below the 1~kcal/mol chemical accuracy threshold for the vast majority of systems, with the few remaining outliers confined almost exclusively to the cc-pVQZ basis set. 
LS-RI-Becke and LS-RI-QRCP achieve the lowest errors overall, with median absolute deviations below 0.01~kcal/mol and maximum errors strictly below 0.1~kcal/mol across all basis sets.
Notably, QRCP-based grid pruning attained comparable fidelity to the unpruned Becke grid while reducing the number of grid points by up to 80\% in cc-pVDZ and approximately 50\% in cc-pVQZ (compare Fig.~\ref{fig:grid_sizes_per_atom}).
Cholesky-based pruning achieved an even stronger compression by retaining only $\sim$10\% and $\sim$25\% of grid points of the parent grid in the cc-pVDZ and cc-pVQZ bases, respectively.
Note that the sparsity levels for Cholesky and K-Means are comparable to the ones reported in Refs.~\citenum{Matthews2020ImprovedGrid} and~\citenum{Lee2020SystematicallyImprovable}.
LS-RI-Cholesky and LS-snRI-Cholesky yielded median errors roughly one order of magnitude larger than LS-RI-[Becke/QRCP], yet no outlier exceeded the 1~kcal/mol threshold.
Replacing the full RI fitting target with the snRI approximation introduced only a marginal accuracy penalty, increasing the mean absolute deviation by approximately 0.015~kcal/mol in cc-pVQZ (Tab.~\ref{tab:energy_errors}). 
LS-RI-KMeans produced a qualitatively different error profile: while median errors were lower than those of the Cholesky-based methods, large outliers exceeding 1~kcal/mol appeared in the cc-pVTZ and cc-pVQZ basis sets. 
This failure mode can be attributed to the breakdown of basis-function density as a proxy for the pair density in systems with more complex electronic structure, in which the heuristic clustering criterion is no longer guaranteed to sample the relevant regions of space.
However, the error can be significantly reduced by increasing the number of IPs per auxiliary basis function (from $3.5$ to $6$), as shown in the SI (Sec.~II; Tab.~II) for the W4-11.129 reaction.
Finally, LS-Aux-Becke, despite exhibiting the highest median error of all methods, still kept all errors within the 1~kcal/mol threshold, demonstrating that the auxiliary Coulomb metric captures the ERI features relevant to the MP2 energy with sufficient fidelity,
a result that has not been demonstrated previously.

Turning to the THC-SCF calculations, the accuracy differences among the methods become less pronounced. 
The four methods LS-RI-Becke, LS-RI-QRCP, LS-RI-Cholesky, and LS-snRI-Cholesky all yield closely grouped mean and median absolute deviations, with the MAD varying by only 0.01~kcal/mol in cc-pVQZ (Tab.~\ref{tab:energy_errors}). 
Cholesky- and QRCP-based pruning still reduced the cc-pVDZ grid by approximately 75\%, yet the cc-pVQZ basis required, on average, more than 50\% of the parent Becke grid points to maintain comparable accuracy.
This illustrates that the AO-ERI is a more complex fitting target.
The snRI fitting target reproduced the LS-RI-Cholesky accuracy closely across all basis sets. 
LS-RI-KMeans produced systematically higher errors, with MADs approximately 0.08~kcal/mol above those of LS-RI-Becke in cc-pVQZ.
We investigated the influence of the chosen parameters on large outliers in the cc-pVQZ basis.
As can be seen from Tab.~II in the SI, the deviation can be reduced by either increasing the parent grid size (from level 0 to level 1) or by increasing the number of IPs per auxiliary basis function in the case of K-Means grid pruning.

LS-Aux-Becke showed substantially larger errors than all RI- and snRI-based methods: Between 31 and 34 systems exceeded the 1~kcal/mol threshold for each of our basis sets.
The most extreme outlier, DC13.13 (electrophilic aromatic substitution of benzene by Cl$_2$), deviated from the RI reference by 12.43~kcal/mol in cc-pVDZ.
This is the largest deviation observed across the entire dataset. 
As demonstrated in the SI (Sec.~II; Tab.~II), these deviations can be systematically reduced by employing a larger auxiliary fitting basis.

Overall, we also observed that for all LS-RI methods larger basis sets yielded higher average errors, reflecting the increased difficulty of representing the ERI in extended basis sets. Furthermore, all \texttt{PyTHC} methods fitting the full RI-ERI or its semi-numerical approximation (snRI) faithfully reproduced the reaction energy differences with errors remaining at or below the 1~kcal/mol threshold for the vast majority of systems. For the rare outliers, we demonstrate in the SI that increasing the grid density converges these errors below the threshold.

\subsection{Timings}
\label{sec:timings}

\begin{figure*}[tbp]
    \includegraphics[width=\linewidth]{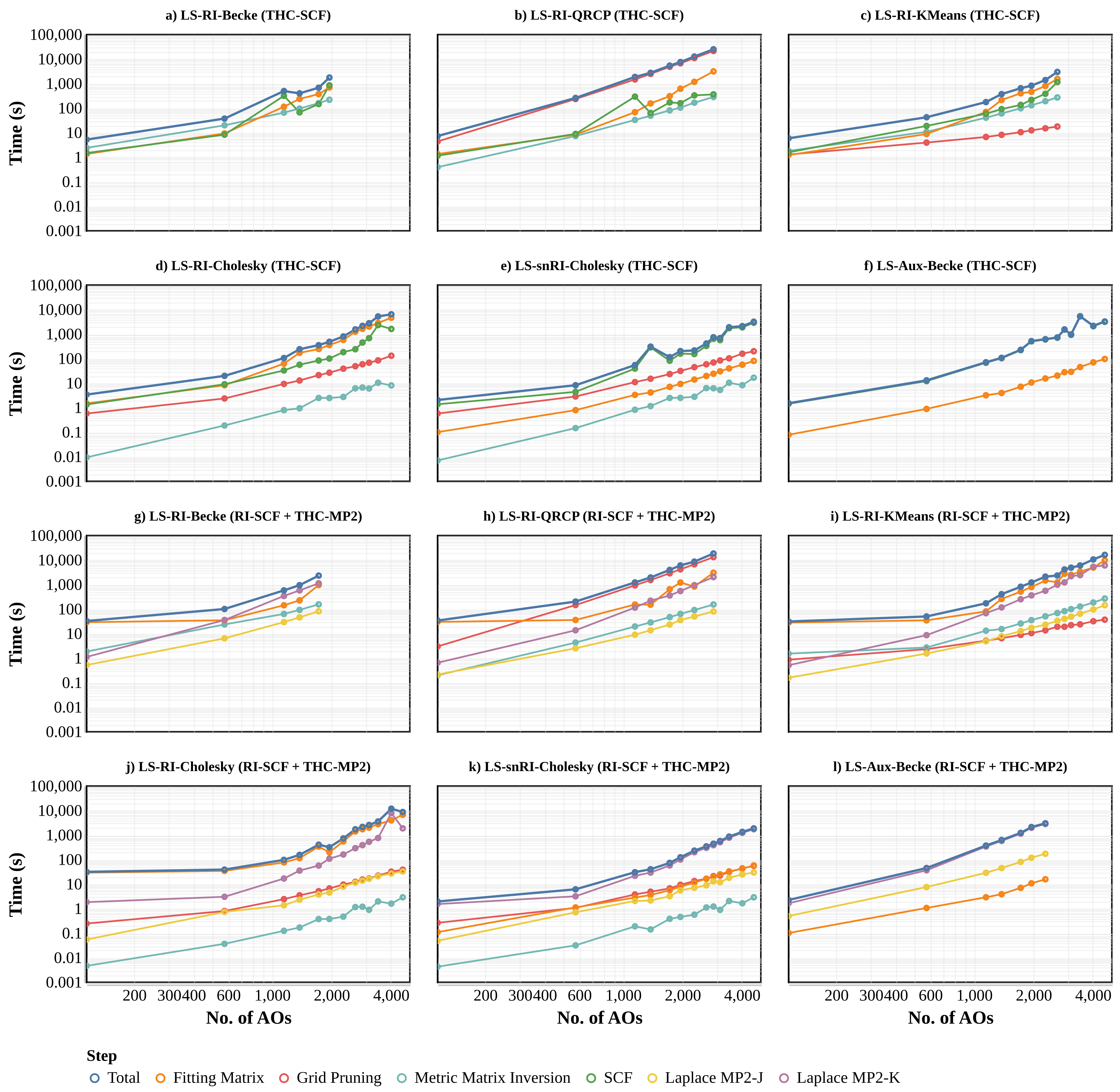}
    \caption{Wall-clock times of the individual algorithmic steps in the THC-SCF and RI-SCF + THC-MP2 calculations on linearly growing \hhon~clusters in the cc-pVQZ basis set. The largest system ($n=40$) contains 4600 basis functions.}
    \label{fig:timings}
\end{figure*}

\begin{figure*}[tpb]
    \includegraphics[width=.85\linewidth]{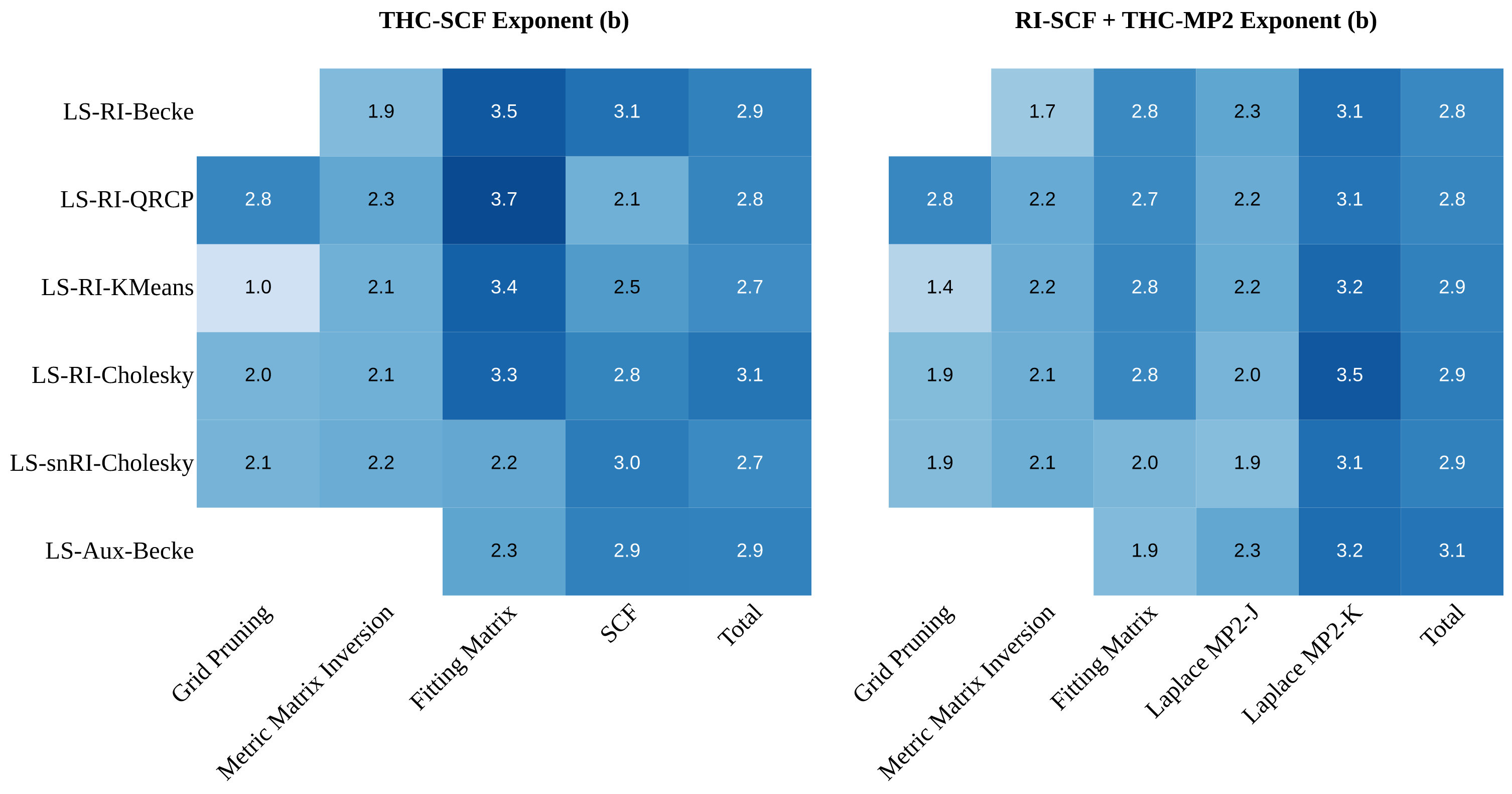}
    \caption{Empirical scaling exponents $b$ for each algorithmic step of the LS-THC methods, determined by fitting the model $t = \alpha \cdot (N_{\mathrm{AO}})^b$ to all data points with $N_{\mathrm{AO}} > 500$ via linear regression on a log--log scale: a) THC-SCF, b) RI-SCF + THC-MP2.}
    \label{fig:scaling_exponents}
\end{figure*}

\begin{table*}
\renewcommand{\arraystretch}{1.2}
\caption{Measured wall-clock execution times (in seconds) for individual algorithmic steps at $N_{\mathrm{AO}} = 1725$ [(\hho)$_{15}$~in the cc-pVQZ basis].}
\label{tab:timings_1725}
\begin{tabular}{llccccccc}
\toprule
 &  & Total & Fitting Matrix & Grid Pruning & Metric Matrix Inversion & SCF & Laplace MP2-J & Laplace MP2-K \\
Type & Method &  &  &  &  &  &  &  \\
\midrule
\multirow[t]{6}{*}{THC-SCF} & LS-RI-Becke & 685 & 376 & - & 162 & 147 & - & - \\
 & LS-RI-QRCP & 5450 & 309 & 4887 & 82 & 172 & - & - \\
 & LS-RI-KMeans & 653 & 406 & 11 & 100 & 136 & - & - \\
 & LS-RI-Cholesky & 357 & 248 & 21 & 3 & 85 & - & - \\
 & LS-snRI-Cholesky & 116 & 7 & 24 & 3 & 83 & - & - \\
 & LS-Aux-Becke & 231 & 7 & - & - & 224 & - & - \\
\cline{1-9}
\multirow[t]{6}{*}{RI-SCF + THC-MP2} & LS-RI-Becke & 2342 & 965 & - & 157 & - & 83 & 1137 \\
 & LS-RI-QRCP & 3999 & 649 & 2916 & 48 & - & 24 & 362 \\
 & LS-RI-KMeans & 823 & 520 & 9 & 26 & - & 13 & 255 \\
 & LS-RI-Cholesky & 409 & 342 & 5 & 0 & - & 4 & 58 \\
 & LS-snRI-Cholesky & 74 & 6 & 7 & 0 & - & 3 & 58 \\
 & LS-Aux-Becke & 1240 & 7 & - & - & - & 81 & 1151 \\
\cline{1-9}
\bottomrule
\end{tabular}
\end{table*}

Having compared the accuracy and grid sizes of the various LS-THC procedures, we assess their practical computational complexity next.
For this, we performed HF-SCF and MP2 calculations for a series of increasingly large water clusters, \hhon~(up to $n=40$), in cc-pVQZ basis.
This means we include systems up to 4600 basis functions in our assessment.
The resulting wall-clock times are displayed in Fig.~\ref{fig:timings}, and the
corresponding empirical scaling exponents for each sub-step are summarized in Fig.~\ref{fig:scaling_exponents}. 
To provide a tangible runtime comparison, Tab.~\ref{tab:timings_1725} reports measured wall-clock timings for (\hho)$_{15}$~($N_{\mathrm{AO}} = 1725$), the largest system completed by all methods.
In all cases, we differentiate between the individual algorithmic steps of the THC construction, including grid pruning, metric matrix inversion (cf.\ Eqs.~\eqref{eq:SMat} and~\eqref{eq:SMat2}), and fitting target construction [for example: Eq.~\eqref{eq:ZDDT}], as well as the subsequent HF-SCF and MP2-J/K calculations.

Regarding grid pruning and metric matrix inversion, the choice of strategy has a decisive impact on the initial computational overhead. 
The unpruned Becke grid and the QRCP algorithm are substantially slower than K-Means or Cholesky-based approaches [Fig.~\ref{fig:timings} a), b), g), h)].
The large grid in LS-RI-Becke increases the computational cost both in the THC and in the subsequent SCF and MP2 calculations.
While QRCP-based pruning effectively reduces the grid size, the pruning step itself [red lines in Fig.~\ref{fig:timings} b), h)] carries a severe computational overhead.
This is why we conducted calculations only for clusters up to (\hho)$_{15}$~and (\hho)$_{25}$~for LS-RI-Becke and LS-RI-QRCP, respectively.

The computational overhead of the QRCP procedure is a direct consequence of the expensive column-norm updates required by the standard Householder-based algorithm, even when combined with randomized matrix compression, as can be seen in Tab.~\ref{tab:timings_1725}.
In contrast, K-Means clustering exhibited near-linear empirical scaling and operated at speeds comparable to or faster than Cholesky pruning [Fig.~\ref{fig:timings} c), i)].
However, the resulting pruned grid can still be relatively large, and the subsequent metric matrix pseudoinversion can contribute significant runtime overhead since it is not guaranteed that the metric matrix is non-singular.
Cholesky decomposition, on the other hand, enables the inversion via highly efficient triangular solvers rather than a full eigenvalue decomposition [Eqs.~\eqref{eq:cd1} and \eqref{eq:cd2}].\cite{Matthews2020ImprovedGrid}
This reduces the inversion cost by up to two orders of magnitude. 
Overall, Cholesky decomposition yields the fastest grid pruning in the MP2 case and the fastest metric matrix inversion across all methods [Fig.~\ref{fig:timings} d), j)], without reaching its formal $\mathcal{O}(N^3)$ scaling on the level-0 parent grid (Fig.~\ref{fig:scaling_exponents}).

Considering the fitting target, the standard RI-ERI fitting is formally the most expensive step for LS-RI-[*] methods.
This is also evident in the scaling exponents in Fig.~\ref{fig:scaling_exponents}.
Its cost is driven primarily by the $\mathcal{O}(N^4)$ contraction of the three-center integrals with the co-density matrix [Eq.~\eqref{eq:ls_thc_df_fit}].
The fitting time is, however, proportional to the grid size. Using the sparser Cholesky grid reduced the fitting time by almost 60\% relative to the full Becke grid in our reference system.

Employing the snRI or auxiliary Coulomb fitting targets formally reduces the fitting step to $\mathcal{O}(N^3)$ complexity, with immediate benefits for the measured wall-clock times [Fig.~\ref{fig:timings} e), f), k), l)]. 
With snRI fitting, the time required to construct the fitting matrix (orange line) drops to nearly the same cost as the grid pruning step (red line). 
The auxiliary Coulomb fitting is still faster, as it entirely circumvents both the three-center integral contraction and the metric matrix inversion. For both of these lower-scaling targets, the THC construction is no longer the dominant step; the overall runtime is instead governed by the downstream electronic structure calculation. 
For the systems considered here, neither method fully reached its formal $\mathcal{O}(N^3)$ asymptote, yielding empirical exponents in the quadratic to sub-quadratic regime.

For the SCF calculations, the wall-clock times exhibit noticeable fluctuations across the water clusters [green lines in Fig.~\ref{fig:timings} a)-e)], due to the different number of SCF iterations for the different water clusters.
However, we find that the empirically determined scaling exponents for the SCF remained below or at the expected asymptotic scaling of $\mathcal{O}(N^3)$.
The measured times at the reference system size (Tab.~\ref{tab:timings_1725}) show large differences between different grid-pruning strategies, reflecting the strong dependence of the THC-ERI contraction on the number of retained grid points.

The MP2 calculations exhibit clearer scaling trends [Fig.~\ref{fig:timings} g)-l)].
The $J^{\text{MP2}}$ evaluation (yellow lines) is highly efficient, showing almost no system-to-system fluctuations and empirical scaling strictly below $\mathcal{O}(N^3)$ across all methods. 
The $K^{\text{MP2}}$ contraction (purple lines) is substantially more expensive, being up to one order of magnitude slower than the $J^{\text{MP2}}$ step in the reference system, with empirical exponents reaching $b \approx 3.5$.
These scalings are higher by approximately one order in $N$ relative to the Coulomb step, in agreement with the theoretical expectation. 
Since both contractions involve the THC grid index multiple times, their cost scales strongly with the retained grid size. This is illustrated in Tab.~\ref{tab:timings_1725}: The wall-time advantage gained by the fast LS-Aux-Becke fitting over LS-RI-Cholesky is entirely offset by the approximately $20\times$ faster $K^{\text{MP2}}$ contraction achieved by the latter two methods due to their sparser grids. 

\section{Conclusion}
\label{sec:conclusion}

In this work, we investigated various linear least-squares tensor hypercontraction (LS-THC) techniques with respect to their accuracy and computational efficiency.
For this, we developed a new open-source Python library, \texttt{PyTHC}, that provides a unified framework for the THC decomposition of the electron-repulsion integral (ERI) and enables consistent benchmarking across different fitting targets and grid-pruning strategies.
We specifically focused on the performance of the different THC methods for the Hartree--Fock self-consistent field (HF-SCF) and second-order M{\o}ller--Plesset perturbation theory (MP2) methods.
For the first time, \texttt{PyTHC} enabled consistent benchmarking across three distinct fitting targets (full RI-ERI, semi-numerical RI-ERI, and the auxiliary Coulomb metric) and three adaptive grid-pruning strategies (pivoted Cholesky decomposition, QRCP, and K-Means clustering) [compare Fig.~\ref{fig:thc_family_tree}].

Our benchmarks for the Diet-GMTKN55 dataset demonstrated that THC methods are capable of achieving chemical accuracy (errors below 1 kcal/mol) for both HF-SCF and MP2 calculations, even for challenging reactions containing up to 81 atoms, heavy elements up to iodine, and large basis sets (up to quadruple-zeta quality). 
Furthermore, we were able to deduce robust default parameters for different THC methods and distill advantages and limitations of each approach.
The overall highest accuracy is achieved for the original LS-THC method (here in conjunction with Becke-type grids; LS-RI-Becke), and the LS-RI-QRCP method, which employs a pivoted QR decomposition to prune the parent grid.
While the QRCP procedure heavily reduces the number of grid points, the high computational cost of the pruning procedure itself in combination with the $\mathcal{O}(N^4)$ fitting step in the case of the RI approximation makes this approach impractical for larger systems.

As an alternative, we showed that K-Means clustering provides a linear-scaling grid-pruning strategy (denoted as LS-RI-KMeans) that achieves competitive grid compression ratios and accurate results for the majority of systems. However, the K-Means approach is sensitive to the choice of the underlying weighting function [weighting function used in this work: Eq.~\eqref{eq:weighting_function}], which can lead to suboptimal grid selection for certain systems. Here, Cholesky Decomposition-based pruning (LS-RI-Cholesky) provides a more robust alternative, yielding the most compact grids while still achieving sub-1~kcal/mol accuracy for the vast majority of systems. Furthermore, highly efficient triangular solvers can be employed for the pseudoinversion step in this case, considerably reducing the computational overhead of this step.

To reduce the formal $\mathcal{O}(N^4)$ fitting cost of the RI-ERI target, we investigated the semi-numerical RI (snRI) fitting target (LS-snRI-Cholesky), which reduces the formal scaling of the overall THC procedure to $\mathcal{O}(N^3)$ while incurring only a marginal accuracy penalty. We therefore conclude that LS-snRI-Cholesky provides a good balance between computational efficiency and accuracy.

We also investigated the auxiliary Coulomb metric fitting target (LS-Aux-Becke), which reduces the formal scaling of the fitting step to $\mathcal{O}(N^3)$.\cite{Hillers-Bendtsen2025LoweringScaling,Hillers-Bendtsen2025AcceleratingHartree} However, because grid pruning strategies are numerically less stable in this case, the underlying large Becke grid used in this work made subsequent electronic structure calculations prohibitively expensive. Nevertheless, errors in MP2 calculations are well below the 1~kcal/mol threshold for all reactions in the Diet-GMTKN55 dataset, even for the largest cc-pVQZ basis set when combined with carefully selected auxiliary basis sets.

Ultimately, these results establish \texttt{PyTHC} as a versatile reference platform for future advancements in LS-THC. In particular, we plan to extend the framework to systems with periodic boundary conditions to assess whether the conclusions of this study generalize.\cite{Sharma2022FastExchange,Rettig2023EvenFaster,Smyser2024UseMultigrids,Yang2026InitioManybody} Furthermore, we will systematically investigate ways to exploit additional sparsity either in the THC construction itself or in subsequent electronic structure calculations to apply THC-based techniques for even larger systems.

%%%%%%%%%%%%%%%%%%%%%%%%
\begin{acknowledgments}
J.~T.~acknowledges funding from the Fonds der Chemischen Industrie (FCI) via a Liebig fellowship and support by the Cluster of Excellence ``CUI: Advanced Imaging of Matter'' of the Deutsche  Forschungsgemeinschaft (DFG) (EXC 2056, funding ID 390715994).
This work was funded by the Deutsche Forschungsgemeinschaft (DFG, German Research Foundation) - 577598009.
For this work, the HPC-cluster Hummel-2 at the University of Hamburg was used.
The cluster was funded by Deutsche Forschungsgemeinschaft (DFG, German Research Foundation) - 498394658.
J.~T. gratefully acknowledges the generous support by Carmen Herrmann at the University of Hamburg.
We would also like to thank Pierre-François Loos for fruitful discussions and for suggesting the use of the Diet-GMTKN55 benchmark set.
\end{acknowledgments}
%%%%%%%%%%%%%%%%%%%%%%%%

%%%%%%%%%%%%%%%%%%%%%%%%%%%%%%%%
\section*{Data availability statement}
%%%%%%%%%%%%%%%%%%%%%%%%%%%%%%%%

The \texttt{PyTHC} library is available at \url{https://github.com/QuantumCorrelators/pythc}.
The data that supports the findings of this study are available within the article and its supplementary material. The raw data is available at \textsc{Zenodo} under the DOI: \url{https://doi.org/10.5281/zenodo.21902340}.

%%%%%%%%%%%%%%%%%%%%%%%%%%%%%%%%
\section*{References}
%%%%%%%%%%%%%%%%%%%%%%%%%%%%%

%

\end{document}